\documentclass{article}

\usepackage{xcolor}
\definecolor{keywordcolor}{rgb}{0.7, 0.1, 0.1}   % red
\definecolor{tacticcolor}{rgb}{0.0, 0.1, 0.6}    % blue
\definecolor{commentcolor}{rgb}{0.4, 0.4, 0.4}   % grey
\definecolor{symbolcolor}{rgb}{0.0, 0.1, 0.6}    % blue
\definecolor{sortcolor}{rgb}{0.1, 0.5, 0.1}      % green
\definecolor{attributecolor}{rgb}{0.7, 0.1, 0.1} % red

\usepackage{listings}

\usepackage{amsmath, amssymb, upgreek}
\usepackage{bm}
\usepackage{unixode}
\usepackage{url}
\usepackage{todonotes}

\newcommand{\ttt}[1]{\texttt{#1}}

\newcommand{\Type}{\mathrm{Type}}

\newcommand{\uPi}{\Uppi}
\newcommand{\uSigma}{\Sigma}
\newcommand{\lam}{\lambda}
\newcommand{\lx}{\lam x}

\newcommand{\lean}{Lean}
\newcommand{\ueq}{\approx} % \overset{?}{=} % \cong
\newcommand{\ucnstr}[3]{\langle #1 \ueq #2,\ #3\rangle}
\newcommand{\join}[2]{#1 \bowtie #2}
\newcommand{\bchoice}[2]{#1~\ttt{in}~#2}
\newcommand{\jchoice}[3]{\langle #1~\ttt{in}~#2, #3\rangle}
\newcommand{\bif}{\textbf{if}}
\newcommand{\bthen}{\textbf{then}}
\newcommand{\belse}{\textbf{else}}
\newcommand{\belseif}{\textbf{else if}}
\newcommand{\bwhen}{\textbf{when}}

\newcommand{\bwhile}{\textbf{while}}
\newcommand{\breturn}{\textbf{return}}

\newcommand{\blet}{\textbf{let}}
\newcommand{\bin}{\textbf{in}}

\begin{document}

\title{Elaboration in Dependent Type Theory}

\author{Leonardo de Moura, Jeremy Avigad, Soonho Kong,\\ and Cody
  Roux\footnote{Avigad's work partially supported by AFOSR
    FA9550-12-1-0370 and FA9550-15-1-0053.}}

%\author{Leonardo de Moura\inst{1} \and Jeremy Avigad\thanks{Work partially supported by AFOSR FA9550-12-1-0370 and FA9550-15-1-0053.}\inst{2} \and
%  Soonho Kong\inst{3} \and Cody Roux\inst{4}}
% \authorrunning{de Moura, Avigad, Kong, and Roux}
% 
% \institute{Microsoft Research, Redmond \and Departments of Philosophy
%   and Mathematical Sciences, Carnegie Mellon University \and
%   Department of Computer Science, Carnegie Mellon University \and
%   Draper Laboratories}

\maketitle

\begin{abstract}
  To be usable in practice, interactive theorem provers need to
  provide convenient and efficient means of writing expressions,
  definitions, and proofs. This involves inferring information that is
  often left implicit in an ordinary mathematical text, and resolving
  ambiguities in mathematical expressions. We refer to the process of
  passing from a quasi-formal and partially-specified expression to a
  completely precise formal one as \emph{elaboration}.  We describe an
  elaboration algorithm for dependent type theory that has been
  implemented in the \lean{} theorem prover. \lean{}'s elaborator
  supports higher-order unification, type class inference, ad hoc
  overloading, insertion of coercions, the use of tactics, and the
  computational reduction of terms. The interactions between these
  components are subtle and complex, and the elaboration algorithm has
  been carefully designed to balance efficiency and usability. We
  describe the central design goals, and the means by which they are
  achieved.
\end{abstract}

\section{Introduction}

Just as programming languages run the spectrum from untyped languages
like Lisp to strongly-typed functional programming languages like
Haskell and ML, foundational systems for mathematics exhibit a range
of diversity, from the untyped language of set theory to simple type
theory and various versions of dependent type theory. Having a
strongly typed language allows the user to convey the intent of an
expression more compactly and efficiently, since a good deal of
information can be inferred from type constraints. Moreover, a type
discipline catches routine errors quickly and flags them in
informative ways. But there is a downside: as we increasingly rely on
types to serve our needs, the computational support that is needed to
make sense of expressions in efficient and predictable ways becomes
increasingly subtle and complex.

Our goal here is to describe the elaboration algorithm used in a new
interactive theorem prover, \emph{\lean{}}~\cite{Lean, tpl}.  \lean{}
is based on an expressive dependent type theory, allowing us to use a
single language to define datatypes, mathematical objects, and
functions, and also to express assertions and write proofs, in
accordance with the propositions-as-types paradigm. Thus, filling in
the details of a function definition and ensuring it is type correct
is no different from filling in the details of a proof and checking
that it establishes the desired conclusion.

The elaboration algorithm that we describe employs nonchronological
backtracking, as well as heuristics for unfolding defined constants
that are very effective in practice. Elaboration algorithms for
dependent type theory are not well documented in the literature,
making the practice something of a dark art. We therefore also hope to
fill an expository gap, by describing the problem clearly and
presenting one solution.

\lean{}'s elaborator is quite powerful. It solves not only first-order
unification problems, but nontrivial higher-order unification problems
as well. It supports the computational interpretation of terms in
dependent type theory, reducing expressions as necessary in the
elaboration process. It supports ad hoc overloading of constants, and
it can insert coercions automatically. It supports a mechanism for
type class inference in a manner that is integrated with the other
components of the elaboration procedure. It also supports interaction
with built-in and user-defined tactics. The interaction between the
components we have just enumerated is subtle and complex, and many
pragmatic design choices were made to attain efficience and usability.

We start in Section~\ref{section:task} with an overview of the task of
the elaborator, focusing on its outward effects. In other words, we
try to convey a sense of what the elaborator is supposed to do. In
Section~\ref{section:procedure}, we explain the algorithm that is used
to achieve the desired results. We describe related work and draw some
conclusions in Section~\ref{section:conclusions}. Lean is an
open-source project, and the source code is freely available
online.\footnote{\url{http://leanprover.github.io}} Many of the
features discussed below are described in greater detail in a tutorial
introduction to Lean \cite{tpl}.

\section{The elaboration task}
\label{section:task}

What makes dependent type theory ``dependent'' is that types can
depend on elements of other types. Within the language, types
themselves are terms, and a function can return a type just as another
function may return a natural number. \lean{}'s standard library is
based on a version of the \emph{Calculus of Inductive Constructions
  with Universes}~\cite{coc,paulin1993inductive,luo1989}, as are
formal developments in Coq~\cite{coq} and Matita~\cite{matita}. We
assume some familiarity with dependent type theory, and provide only a
brief overview here.

Lean's core syntax is based on a sequence of non-cumulative type
universes and $\uPi$-types. There is an infinite sequence of type
universes \ttt{Type₀}, \ttt{Type₁}, \ttt{Type₂}, \ldots, and any term
\ttt{t~:~Type}$_{\mathrm i}$ is intended to denote a type in the
\ttt{i}th universe. Each universe is closed under the formation of
$\uPi$-types \ttt{$\uPi$x~:~A, B}, where \ttt{A} and \ttt{B} are
type-valued expressions, and \ttt{B} can depend on \ttt{x}. The idea
is that \ttt{$\uPi$x~:~A, B} denotes the type of functions \ttt{f}
that map any element \ttt{a~:~A} to an element of \ttt{B[a/x]}. When
\ttt{x} does not appear in \ttt{B}, the type \ttt{$\uPi$x~:~A, B} is
written \ttt{A → B} and denotes the usual non-dependent function
space.

Lean's kernel can be instantiated in different ways. In the standard
mode, \ttt{Type₀} is distinguished by the fact that it is
\emph{impredicative}, which is to say, \ttt{$\uPi$x~:~A, B} is an
element of \ttt{Prop} for every \ttt{B~:~Prop} and every
\ttt{A}. Elements of \ttt{Prop} are intended to denote propositions.
When \ttt{$\uPi$x~:~A, B} is in \ttt{Prop}, it can also be written as
\ttt{$\forall$x~:~A, B}, and it is interpreted as the proposition that
\ttt{B} holds for every element \ttt{x} of \ttt{A}. Given
\ttt{P~:~Prop}, an element \ttt{t~:~P} can be interpreted as a proof
of \ttt{P}, under the propositions-as-types correspondence, or more
simply as ``evidence'' that \ttt{P} holds. Such a type \ttt{P} is
\emph{proof-irrelevant}, which is to say, any two terms \ttt{s, t~:~P}
are treated as definitionally equal by the kernel.

Lean also provides a predicative mode for homotopy type theory,
without any special treatment of \ttt{Type₀}. The result is a version
of Martin-L\"of type theory~\cite{mltt,hottbook} similar to the one
used in Agda~\cite{agda}.

In both standard and hott modes, the type universes are
non-cumulative. They are treated polymorphically, which is to say,
there are explicit quantifications over universes.  In practice, users
generally write $t~:~\Type$, leaving it to \lean{} to insert an
implicit universe variable and manage universe constraints
accordingly.

Extensions to the core type theory inhabit a second layer of the
kernel. Both modes allow one to form \emph{inductive
  families}~\cite{Dybjer94}, a mechanism that can be used to define
basic types like \ttt{nat} and \ttt{bool}, and common type-forming
operations, like Cartesian products, lists, $\uSigma$-types, and so
on.  Each inductive family declaration generates a \emph{recursor}
(also known as the \emph{eliminator}). The standard mode includes a
mechanism for forming quotient types, and the mode for homotopy type
theory includes certain higher-inductive types \cite{van:doorn:15}. We
need not be concerned with the precise details here, except to note
that terms in dependent type theory come with a computational
interpretation. For example, given \ttt{t~:~B} possibly depending on a
variable \ttt{x~:~A} and \ttt{s~:~A}, the term \ttt{($\lambda$x,
  t) s} reduces to \ttt{t [s / x]}, the result of replacing \ttt{x}
by \ttt{s} in \ttt{t}. Similarly, inductive types support definition
by recursion; for example, if one defines addition on the natural
numbers by structural recursion on the second argument, \ttt{t + 0}
reduces to \ttt{t}. The kernel type checker should identify terms that
are equivalent under the induced equivalence relation, and, as much as
possible, the elaborator should take this equivalence into account
when inferring missing information. We discuss this further in
Section~\ref{subsection:computational}.

\subsection{Type inference and implicit arguments}

The task of the elaborator, put simply, is to convert a partially
specified expression into a fully specified, type-correct term.  For
example, in \lean{}, one can define a function \verb=do_twice= as follows:
\begin{lstlisting}
  definition do_twice (f : ℕ → ℕ) (x : ℕ) : ℕ := f (f x)
\end{lstlisting}
One can omit any two of the three type annotations, leaving it to the
elaborator to infer the missing information. Inferring types like
these can be seen as a generalization of the Hindley-Milner algorithm
\cite{hindley:69,milner:78}, which takes an unsigned term in the
$\lam$-calculus and assigns a simple type to it, if one exists.

% Introduced in the '70s, the Hindley-Milner algorithm takes unsigned
% terms in the $\lam$-calculus and assigns a simple type to it, if one
% exists. It is complete in the sense that it always finds the most
% general possible simple type of a given untyped term.

% The algorithm has gained widespread usage in most mainstream
% programming languages, relieving the burden of annotation for the
% programmer with a good trade-off of power vs speed.

% One possible view of the algorithm is to insert explicit type
% annotations into untyped term, giving types to abstracted variables
% and types which satisfy the typing constraints given by the term
% itself. Indeed, the GHC compiler takes exactly this view, using a
% fully explicitly typed calculus as an intermediate language. This has
% the strong advantage of admitting a complex algorithm for type
% inference which is ``sanity checked'' by a simpler algorithm for the
% fully explicit language.

% Over the years, the Hindley-Milner algorithm has been extended to
% support increasingly powerful type systems, for example higher-kinded
% polymorphism and type classes in Haskell.

When entering a term, a user can leave an argument implicit by
inserting an underscore, leaving it to the elaborator to infer a
suitable value. One can also mark function arguments as implicit by
declaring them using curly brackets when defining the function, to
indicate that they should be inferred rather than entered
explicitly. For example, we can define the \ttt{append} function,
which concatenates two lists, so that it has the following type:
\begin{lstlisting}
  append : Π {A : Type}, list A → list A → list A
\end{lstlisting}
Users can then write \ttt{append l₁ l₂} rather than \ttt{append A l₁
  l₂}, leaving Lean to infer the first argument.

\subsection{Higher-order unification}

Many of the constraint problems that arise in type inference and the
synthesis of implicit arguments are easily solved using first-order
unification. For example, suppose a user writes \ttt{append l₁ l₂},
where \ttt{l₁} can be seen to have type \ttt{list T}. Temporarily naming
the implicit argument to \ttt{append} as \ttt{?M}, we see that next
argument to \ttt{append} should have type \ttt{list ?M}. Given that \ttt{l₁}
in fact has type \ttt{list T}, we easily infer \ttt{?M = T}.

Nevertheless, it is often the case that the elaborator is required to
infer an element of a $\uPi$-type, which constitutes a
\emph{higher-order unification} problem. For example, if \ttt{e~:~a =
  b} is a proof of the equality of two terms of some type \ttt{A}, and
\ttt{H~:~P} is a proof of some expression involving \ttt{a}, then the
term \ttt{subst e H} denotes a proof of the result of replacing some
or all of the occurrences of \ttt{a} in \ttt{P} with \ttt{b}. Here, in
addition to inferring the type \ttt{A}, we also need to infer an
expression \ttt{T~:~A → Prop} denoting the context for the
substitution, that is, the expression with the property that \ttt{T a}
is convertible to \ttt{P}. Such an expression is inherently ambiguous;
for example, if \ttt{H} has type \ttt{R (f a a) a}, then with
\ttt{subst e H} the user may have in mind \ttt{R (f b b) b} or \ttt{R
  (f a b) a} or something else, and the elaborator has to rely on
context and a backtracking search to find an interpretation that
fits. Similar issues arise with proofs by induction, which require the
system to infer an induction predicate.

The need for higher-order unification even arises with common
datatypes. For example, the type \ttt{$\uSigma$x~:~A, B} denotes the
type of dependent pairs \ttt{⟨a, b⟩}, where \ttt{a~:~A} and \ttt{b~:~B
  a}. Here \ttt{B} is in general a function \ttt{A → Type}. In the
notation \ttt{⟨a, b⟩}, the arguments \ttt{A} and \ttt{B} are left
implicit. The argument \ttt{A} can easily be inferred from the type of
\ttt{A}, but the type of \ttt{b} will generally be an expression that
involves \ttt{a} as an argument. In this case, higher-order
unification is used to infer \ttt{B}.

Even second-order unification is known to be generally
undecidable~\cite{goldfarb81}, but the elaborator merely needs to
perform well on instances that come up in practice.
For example, in Lean, users can import the
notation \lstinline{H ▸ H'} for \ttt{subst H H'}. If we have proved
\begin{lstlisting}
  theorem mul_mod_mul_left {z : ℕ} (x y : ℕ) (zpos : z > 0) :
          (z * x) mod (z * y) = z * (x mod y)
\end{lstlisting}
we can then write
\begin{lstlisting}
  theorem mul_mod_mul_right (x z y : ℕ) :
          (x * z) mod (y * z) = (x mod y) * z :=
  !mul.comm ▸ !mul.comm ▸ !mul.comm ▸ !mul_mod_mul_left
\end{lstlisting}
The proof applies the commutativity of multiplication three times to
an appropriate instance of the theorem \verb=mul_mod_mul_left=. (The
symbol \ttt{!} indicates that all arguments should be synthesized by
the elaborator.) The unifier can similarly handle nested inductions
and iterated recursion.
\begin{lstlisting}
  theorem add.comm (n m : ℕ) : n + m = m + n :=
  nat.induction_on m
   (nat.induction_on n rfl
     (take n, assume IH : n = 0 + n,
       show succ n = succ (0 + n), from IH ▸ rfl))
     -- induction step omitted
\end{lstlisting}
Or, in the context of homotopy type theory, where equality proofs are
\emph{relevant}, we can write:
\begin{lstlisting}
  definition concat_assoc (p : x = y) (q : y = z) (r : z = t) :
             p ⬝ (q ⬝ r) = (p ⬝ q) ⬝ r :=
  eq.rec_on r (eq.rec_on q idp)
\end{lstlisting}
Here \verb=rec_on= denotes a form of recursion which, like induction,
has to infer the relevant predicate.

We will see below that higher-order unification is a complex process,
and places a high burden on the elaborator. It should thus be used
sparingly. But it is often convenient and sometimes unavoidable, so it
is important that it can be handled by the elaboration algorithm.

\subsection{Computational behavior}
\label{subsection:computational}

The elaborator should also respect the computational interpretation of
terms. It should, for instance, recognize the equivalence of the terms
\ttt{(λx, t) s} and \ttt{t[s/x]}, as well as \ttt{⟨s, t⟩.1} (denoting
the first projection of the pair) and \ttt{s} under the relevant
reduction rule for pairs. Elements of inductive types also have
computational behavior; on the natural numbers, \ttt{2 + 2} and
\ttt{4} are both definitionally equal to \ttt{succ (succ (succ (succ
  0)))}, \ttt{x + 0} is definitionally equal to \ttt{x}, and \ttt{x +
  1} is definitionally equal to \ttt{succ x}. The elaborator should
also support unfolding definitions where necessary: for example, if
\ttt{x - y} is defined as \ttt{x + (-y)}, the elaborator should allow
us to use the commutativity of addition to rewrite \ttt{x - y} to
\ttt{-y + x}. Unfolding definitions and reducing projections is
especially crucial when working with algebraic structures, where many
basic expressions cannot even be seen to be type correct without
carrying out such reductions. For example, given \ttt{a~:~A} and
\ttt{b~:~B a}, the left-hand side of the expression \ttt{⟨a, b⟩.2 = b}
has type \ttt{B (⟨a, b⟩.1)} and the right-hand side has type \ttt{B
  a}. Both the elaborator and the type checker need to recognize these
types as the same.

Determining when to unfold defined constants is a crucial part of the
practice of theorem proving. It is therefore unsurprising that the
naive approach of performing \emph{all} such unfoldings leads to
unacceptable performance, and it is an important aspect of building a
practical elaboration procedure to design heuristics that limit
unfolding to situations that require it. Lean allows users to annotate
definitions, providing hints to the elaborator, as follows:
\begin{itemize}
\item An \emph{irreducible} definition will never be unfolded during
  higher-order unification (but can still be unfolded in other
  situations, for example during type checking).
\item A \emph{reducible} definition will be always eligible for unfolding.
\item A \emph{semireducible} definition can be unfolded during simple
  decisions and won't be unfolded during complex decisions.
\end{itemize}
For example, users can mark definitions which ought to be viewed as
abbreviations as reducible.  The meaning of these annotations is
discussed further in Section~\ref{subsection:support}. These
annotations are used only by the elaborator; they have no bearing at
all when it comes to checking the type of a fully elaborated term. As
a result, the user can modify these annotations at any time, as
needed, when developing a theory.

\subsection{Type classes}

\lean{} supports the use of Haskell-style \emph{type
  classes}~\cite{hall1996type}. For example, we can define a class
\verb=has_mul A= of types \ttt{A} with an associated multiplication
as follows:
\begin{lstlisting}
  structure has_mul [class] (A : Type) := (mul : A → A → A)
\end{lstlisting}
In other words, for every type \ttt{A}, \verb=has_mul A= is a record
with one element, \verb=has_mul.mul=, which we should think of as a
multiplication operation on \ttt{A}. We then define the generic
multiplication operation,
\begin{lstlisting}
  definition mul {A : Type} [s : has_mul A] : A → A → A :=
  has_mul.mul
\end{lstlisting}
and the notation
\begin{lstlisting}
  infix * := mul
\end{lstlisting}
The square brackets indicate that the argument \ttt{s} is implicit,
and that the relevant instance of \verb=has_mul A= should be
synthesized by the class inference mechanism. We can declare a
particular instance as follows:
\begin{lstlisting}
  definition nat_has_mul [instance] : has_mul nat :=
  has_mul.mk nat.mul
\end{lstlisting}
Suppose the user writes \ttt{s * t}, when \ttt{s} is inferred to have type
\ttt{nat}. When the elaborator is called to solve
\verb=?M : has_mul nat=,
it finds \verb=nat_has_mul= on a stored list of instances, and
assigns that to \ttt{?M}.

Instance declarations themselves can have implicit class arguments, in
which case, class inference performs a backward-chaining Prolog-like
search. For example, we can declare
\begin{lstlisting}
  structure semigroup [class] (A : Type) extends has_mul A :=
  (mul_assoc : ∀a b c, mul (mul a b) c = mul a (mul b c))
\end{lstlisting}
The \ttt{structure} declaration above automatically declares
\ttt{semigroup} to be an instance of \verb=has_mul=, so that if,
instead, \ttt{nat} was only declared to be an instance of
\ttt{semigroup}, class inference could synthesize the instance of
\verb=has_mul nat= in two steps. We then get the generic theorem
\ttt{mul.assoc} in the same way we obtained the generic notation for
multiplication.

We can then go on to define monoids, groups, rings, and commutative
versions. The \ttt{structure} command supports the construction of the
algebraic hierarchy by allowing the user to extend and merge multiple
structures:
\begin{lstlisting}
  structure group [class] (A : Type) 
            extends monoid A, has_inv A :=
  (mul_left_inv : ∀a, mul (inv a) a = one)
\end{lstlisting}
Users can also rename structure components on the fly. In the
following example, type class inference finds the appropriate inverse
and instance of the theorem \verb=inv_inv= when processing \verb=eq_inv_of_eq_inv=:
\begin{lstlisting}
  theorem inv_inv {A : Type} [s : group A] (a : A) : 
          (a⁻¹)⁻¹ = a := 
  ...

  theorem eq_inv_of_eq_inv {A : Type} [s : group A] {a b : A}
                           (H : a = b⁻¹) : b = a⁻¹ :=
  by rewrite [H, inv_inv]
\end{lstlisting}
Here, the \ttt{rewrite} tactic replaces \ttt{a} by
\ttt{b⁻¹} in the goal, and then rewrites \ttt{(b⁻¹)⁻¹} to \ttt{b}. Since any
\ttt{group} is an instance of a \ttt{monoid} and any \ttt{monoid} is
an instance of a \ttt{semigroup}, generic theorems about semigroups and
monoids can be applied to any group.
The type class inference is seamless integrated in all proof procedures
implemented in Lean. We remark that the theorem above can be proved without
any user guidance using these procedures.

We can also declare so called \emph{fully bundled structures} in the
style of the Mathematical Components library~\cite{mathstruct}. For
example:
\begin{lstlisting}
  structure Group := (carrier : Type) (struct : group carrier)

  attribute Group.carrier [coercion]
  attribute Group.struct  [instance]
\end{lstlisting}
This means that whenever we have \ttt{G~:~Group}, we can write
\ttt{g~:~G}, and \ttt{G} is coerced to its carrier type. And whenever we
have \ttt{g~:~carrier G}, type class resolution can infer that
\ttt{struct G} is the relevant group structure on \ttt{carrier G}.

In Lean, type classes can be used to infer not only notation and
generic facts, but fairly complex data. For example, in the standard
library, we define the class of propositions that are decidable:
\begin{lstlisting}
  inductive decidable [class] (p : Prop) : Type :=
  | inl :  p → decidable p
  | inr : ¬p → decidable p
\end{lstlisting}
Logically speaking, since \ttt{decidable p} lives in \ttt{Type} rather
than \ttt{Prop}, having an element \ttt{t~:~decidable p} is more
informative than having an element \ttt{t~:~p ∨ ¬p}; it enables us to
define values of an arbitrary type depending on the truth value of
\ttt{p}. This distinction is only useful in constructive mathematics,
because classically every proposition is decidable. But the
\ttt{decidable} class allows for a smooth transition between
constructive and classical logic, allowing classical reasoning in
suitable constructive settings as well. It is especially relevant
for users interested in defining computable functions.  In Lean,
\ttt{(if c then t else e)} is notation for \ttt{(ite c t e)}, where
\ttt{ite} is defined as:
\begin{lstlisting}
  definition ite (c : Prop) [H : decidable c] {A : Type} 
                 (t e : A) : A :=
  decidable.rec_on H (λ Hc, t) (λ Hnc, e)
\end{lstlisting}
Note that the implicit argument \ttt{H} is automatically synthesized
by type class inference. Moreover, the expression \ttt{if c then t else e}
computes whenever \ttt{c} is a decidable proposition.

For example, we can prove, constructively, that equality on the
natural numbers is decidable:
\begin{lstlisting}
  nat.decidable_eq [instance] : ∀ x y : nat, decidable (x = y)
\end{lstlisting}
We can do the same for inequality relations on \ttt{nat}, and moreover show
that decidability is preserved under boolean operations and bounded
quantification. We moreover make the following definitions:
\begin{lstlisting}
  definition is_true (c : Prop) [H : decidable c] : Prop :=
  if c then true else false

  definition of_is_true {c : Prop} [H₁ : decidable c]
                        (H₂ : is_true c) : c :=
  decidable.rec_on H₁ (λ Hc, Hc)
                      (λ Hnc, !false.rec (if_neg Hnc ▸ H₂))

  notation `dec_trivial` := of_is_true trivial
\end{lstlisting}
What is going on here is subtle. The expression \verb=is_true c=
infers a decision procedure for \ttt{c}, and returns either \ttt{true}
or \ttt{false}. Assuming \verb=H₂: is_true c=, \verb=of_is_true H₂= is
a proof of \ttt{c}. But if \verb=is_true c= evaluates to true, it has
the canonical proof \ttt{trivial}. Thus, given a proposition \ttt{c},
the notation \verb=dec_trivial= does the following:
\begin{itemize}
\item infers a decision procedure for \ttt{c}, and
\item tries to use \ttt{trivial} to prove \verb=is_true c=.
\end{itemize}
If it succeeds --- that is, if the resulting term type checks --- the
result is a proof of \ttt{c}.

With these definitions, we can write the following proof:
\begin{lstlisting}
  example : ∀ x : nat, x < 10 → x ≠ 10 ∧ x < 12 := dec_trivial
\end{lstlisting}
Type class resolution infers the decision procedure for the
proposition in question, and computational reduction evaluates
it. Problems like this can appear anywhere in a proof or an
expression, and type class inference will solve them at appropriate
times within the elaboration process.

\subsection{Overloading}

We have seen that the standard library relies on type class inference
to support the use arithmetic operations like \ttt{+} and \ttt{*} for
different number classes. This is sometimes known as \emph{parametric
  polymorphism}. Lean also supports \emph{ad hoc polymorphism} by
allowing us to overload identifiers and notation. For example, the
notation \ttt{++} is used for concatenation of both lists and tuples:
\begin{lstlisting}
  import data.list data.tuple
  open list tuple

  variables (A : Type) (m n : ℕ)
  variables (v : tuple A m) (w : tuple A n) (s t : list A)

  check s ++ t
  check v ++ w
\end{lstlisting}
Where it is necessary to disambiguate, Lean allows us to precede an
expression with the notation \verb=#<namespace>=, to specify the
namespace in which notation is to be interpreted.
\begin{lstlisting}
  check λ x y, (#list x + y)
  check λ x y, (#tuples x + y)
\end{lstlisting}

We can also overload identifiers. Every identifier in Lean has a full
name that is unique, but identifiers can be grouped into namespaces,
and opening the namespace produces a shorter alias. For example, if we
define \ttt{foo} in namepsaces \ttt{a} and \ttt{b}, we obtain
identifiers named \ttt{a.foo} and \ttt{b.foo} respectively. If we open
both namespaces, however, the alias \ttt{foo} is an overloaded
reference to both, leaving the elaborator to resolve the ambiguity.

Ad hoc overloading is more flexible than type class overloading, in
that the overloaded constants can denote entirely different kinds of
objects.  It adds ambiguity and choice points to the elaboration
process, and should therefore be used sparingly. But it is often
useful, especially when we want to reuse notation for expressions that
do not have the same shape. For example, in the \lean{} standard
library, we use \ttt{⁻¹} above to denote the inverse function for
algebraic structures that support it, as well as for the symmetry
operation for equalities. It can also be used to invert bijections,
used group isomorphisms, ring isomorphisms, isomorphisms in a
category, or equivalences between categories.

\subsection{Coercions}
\label{subsection:coercions}

The treatment of coercions in Lean is as one would expect. One can,
for example, coerce a \ttt{bool} to a \ttt{nat} and a \ttt{nat} to an
\ttt{int}, and Lean will insert coercions in list expressions \ttt{[n,
    i, m, j]} and \ttt{[i, n, j, m]} when \ttt{n} and \ttt{m} have
type \ttt{nat} and \ttt{i} and \ttt{j} have type \ttt{int}. One can
also coerce axiomatic structures, so that the user can provide a group
as input anywhere a semigroup is expected. One can also coerce from a
suitable family of types to \ttt{Type} or to a $\uPi$-type.

In fact, just as in Coq, Lean allows us to declare three kinds of
coercions:
\begin{itemize}
\item from a family of types to another family of types
\item from a family of types to the class of sorts
\item from a family of types to the class of function types
\end{itemize}
The first kind of coercion allows us to view any element of a member
of the source family as an element of a corresponding member of the
target family. The second kind of coercion allows us to view any
element of a member of the source family as a type. The third kind of
coercion allows us to view any element of the source family as a function.
For details, see \cite{tpl}.

\subsection{Tactics and structuring mechanisms}
\label{subsection:tactics}

Finally, definitions and proofs can invoke \emph{tactics}, that is,
user-defined or built-in proof-finding procedures that construct various
subterms. The constraint solver described in this paper invokes user
provided tactics to construct terms that cannot be synthesized by
solving unification constraints and type class resolution.  Lean's
tactic language is similar to those found in other LCF-style theorem
provers. Describing the tactic language here would take us too far
afield; we only wish to point out that our implementation makes the
use of tactics continuous with the act of writing terms. Anywhere a
term is expected, a user can used the keywords \ttt{begin} and
\ttt{end} to enter a tactic block:
\begin{lstlisting}
  theorem test (p q : Prop) (Hp : p) (Hq : q) : p ∧ q ∧ p :=
  begin
    apply and.intro,
    exact Hp,
    apply and.intro,
    exact Hq,
    exact Hp
  end
\end{lstlisting}
One-line tactic proofs can be specified with the keyword \ttt{by}:
\begin{lstlisting}
  theorem test (p q : Prop) (Hp : p) (Hq : q) : p ∧ q ∧ p :=
  by apply (and.intro Hp); exact (and.intro Hq Hp)
\end{lstlisting}
Conversely, when in tactic mode, one can use the \ttt{exact} tactic to
specify an explicit term, as in the example above. The keywords
\ttt{have} and \ttt{show} make it possible to do that in an elegant and
structured way.
\begin{lstlisting}
  theorem card_image_eq_of_inj_on {f : A → B} {s : finset A}
                                  (H1 : inj_on f (ts s)) :
          card (image f s) = card s :=
  begin
    induction s with a t H IH,
     {rewrite [card_empty]},
     {have H2 : ts t ⊆ ts (insert a t),
        by rewrite [-subset_eq_to_set_subset];
          apply subset_insert,
      have H3 : card (image f t) = card t,
        from IH (inj_on_of_inj_on_of_subset H1 H2),
      have H4 : f a ∉ image f t,
        from ..., -- proof suppressed
      show card (image f (insert a t)) = card (insert a t),
        from ... -- proof suppressed}
  end
\end{lstlisting}
Thus one can pass freely between the two modes. This yields a tradeoff
between two different strategies for elaboration: tactics build
an expression using local information in a surgical way, whereas the
elaborator solves constraints involving global information, spread out
across the entire term.

The availability of tactic mode also provides a convenient way of
sectioning long proof terms: the construct \ttt{proof t qed} is
syntactic sugar for \ttt{by+ exact t}. Including this in a long proof
terms forces the elaborator to process the surrounding expression
independent of \ttt{t}, and then process \ttt{t}, separately, using
information from the surrounding term. Thus we can treat the
processing of a long proof as one large elaboration problem or the
composition of smaller ones, balancing the advantages of the local and
global approaches in a convenient and flexible way.

\subsection{Combining the various components}

Any given definition or theorem in Lean can draw on many of
the features just described. Consider the following, which defines the
composition of two natural transformations between functors (and
establishes that it is, indeed, a natural transformation):
\begin{lstlisting}
  variables {C D : Category} {F G H : C ⇒ D}

  definition nt_compose (η : G ⟹ H) (θ : F ⟹ G) : F ⟹ H :=
  natural_transformation.mk
    (take a, η a ∘ θ a)
    (take a b f, calc
      H f ∘ (η a ∘ θ a) = (H f ∘ η a) ∘ θ a : assoc
                    ... = (η b ∘ G f) ∘ θ a : naturality
                    ... = η b ∘ (G f ∘ θ a) : assoc
                    ... = η b ∘ (θ b ∘ F f) : naturality
                    ... = (η b ∘ θ b) ∘ F f : assoc)
\end{lstlisting}
Here the functors \ttt{F}, \ttt{G}, and \ttt{H} are coerced to their
action on morphisms, and the natural transformations \ttt{η} and
\ttt{θ} are coerced to their first component. The composition symbol
\ttt{∘} for functions is overloaded to denote composition of morphisms
as well, and type class inference infers the category in which the
composition takes place. The appropriate substitution contexts in the
calculation are inferred, as are the arguments to the theorems that are
invoked.

The interactions between the components of the elaboration task are
subtle, and the challenge is to deal with them all at the same time. A
definition or proof may give rise to hundreds of constraints requiring
a mixture of higher-order unification, disambiguation of overloaded
symbols, insertion of coercions, type class inference, and
computational reduction. The net effect is then a difficult
constraint-solving problem with a combinatorial explosion of
options. \lean{}'s elaborator manages to solve such problems, and it
is quite fast. (See, for example, the data presented at the end of
Section~\ref{subsection:support}.) In the next section, we explain how
the elaborator processes the constraints and navigates the search
space in an effort to balance completeness and efficiency.

\section{The elaboration procedure}
\label{section:procedure}

\subsection{Overview}

Section~\ref{section:task} describes what we want the elaboration
algorithm to do. It needs to infer types and implicit arguments in
expressions, including sometimes higher-order functions and
predicates. It needs to support type class inference that is robust
enough to work with structures in an algebraic hierachy an a uniform
and convenient way. It needs to dismabiguate overloaded notation and
identifiers, and it needs to insert coercions where
appropriate. Moreover, it needs to respect the computational behavior
of expressions while performing all these tasks, since, in general,
the constraints can only be solved up to equivalence of terms. The
goal of this section is to describe an algorithm that does this.

The task of Lean's parser, which we do not describe here, is to
convert a user's input to a \emph{preterm}, a formal but incomplete
reflection of that input.  The process for getting from a preterm to a
fully elaborated term has two main steps: \emph{preprocessing} and
\emph{constraint resolution}. The preprocessing phase takes a preterm
and creates a partially specified term, with ``holes,'' or
metavariables, representing the information that needs to be
inferred. At the same time, the preprocesor generates a list of
constraints that these metavariables need to satisfy. Some of the
constraints are \emph{unification constraints}, for example, the
constraint that the type of an argument to a function matches the
function's argument type. Others are \emph{choice constraints}, for
example, that an inferred value is among a finite set of possible
overloads. Finally, after all constraints have been solved, Lean
invokes the tactic blocks associated with the remaining holes to
produce the terms necessary to fill them. This is a recursive
procedure because some tactics may contain nested preterms that must
be also elaborated, and these preterms may additional tactic blocks,
and so on.

The constraint resolution phase aims to find a consistent solution
to all the unification and choice constraints. Heuristically,
``straightforward'' constraints should be solved first, providing
useful information to guide the rest of the search. Choice points
result in backtracking, which needs to be handled carefully to avoid
duplication of work. Failures need to be carefully tracked in order to
provide informative error messages to the user.

The simple division into the preprocessing phase and constraint
resolution phase is slightly too simplistic: even the preprocessing
phase has to process and simplify constraints, in order to detect the
possibility of inserting a coercion. Thus both phases make use of a
\emph{constraint simplification} procedure that performs preliminary
reductions.

We spell out the details below. Sections \ref{subsection:main} and
\ref{subsection:support} describe the main data structures and some of
the support functions used by the algorithm, and
Section~\ref{subsection:simplification} describes the constraint
simplification procedure. The preprocessing step is described in
Section~\ref{subsection:preprocessing}, and the constraint solving
procedure, which is both the heart of the elaboration algorithm and
the most complex component, is described in
Sections~\ref{subsection:solving} and \ref{subsection:processing}.

\subsection{Main data structures}
\label{subsection:main}

In this section, we describe the term representation and the
main data structures used in our elaboration procedure.  We assume the
term language is a dependent $\lam$-calculus in which terms are
described by the following grammar:
\[t,s \ = \ \ell \mid x \mid f \mid\ ?m \mid \Type\ u \mid t\ s \mid \lam x : s, t \mid \uPi x : s, t
\]
where
\begin{itemize}
\item $\ell$ a free variable (also called a local constant)
\item $x$ is a bound variable
\item $f$ is a constant (parametrized by a list of universe terms)
\item $?m$ is a metavariable
\item $u$ is a universe term
\end{itemize}

We adopt a \emph{locally nameless} variable binding style: free
variables have a unique identifier and a type, while bound variables
are simply represented by a number, a de Bruijn index. We store the
type with each free variable, thereby removing the need to carry
around contexts in the type checker and normalizer.  As described
in~\cite{IamNotANumber}, this representation style simplifies the
implementation considerably, as it minimizes the number of
places where explicit calculations with de Bruijn indices must be performed.
We use the notation $t[x := s]$ to represent the substitution of
$x$ for $s$ in $t$, where $x$ is a bound variable, free variable, or
metavariable. When $x$ is a bound variable, the operation
also lowers all bound variables with index greater than $x$.
We use $\bm{t}$ to denote sequence of terms $t_1 \ldots t_n$,
and $\bm{x} : \bm{A}$ for the telescope $(x_1 : A_1)\, \ldots\, (x_n : A_n)$,
where $A_i$ may depend on $x_j$ for $j < i$.

While the locally nameless approach simplifies many aspects of the
code, the operations of abstracting and instantiating variables can be
costly. Fortunately, we found a simple optimization that completely
eliminates any performance concerns. The problem, and our solution, are
described at the end of Section~\ref{subsection:support}.

An \emph{environment} stores a sequence of declarations.  The \lean{}
kernel supports three different kinds of declarations: \emph{axioms},
\emph{definitions} and \emph{inductive families}. Each has a unique
identifier, and can be parametrized by a sequence of universe
parameters. Every axiom has a type, and every definition has a type
and a value. A constant is just a reference to a declaration.

%% We say a \emph{preterm} is a \lean{} term produced by the parser.  It
%% represents partial constructions provided by users.  A preterm may
%% contain \emph{holes}, missing pieces, that must be synthesized by the
%% system.  The elaborator may also have to introduce coercions to make
%% sure the term can be type checked by the kernel.

A user's input to the elaborator can generally be viewed as
\emph{partial constructions}, i.e., constructions containing
\emph{holes} that must be filled by the system.  Internally, each hole
is represented by a metavariable.  Each metavariable has a unique
identifier and a type. The main operation on metavariables is
\emph{instantiation}.  In our implementation, only closed terms can be
assigned to metavariables. This design decision guarantees that
operations such as $\beta$-reduction and metavariable instantiation
commute.  Since only closed terms can be assigned to metavariables, on
creation a metavariable is applied to the variables in the context
where it appears.  For example, we encode a hole in the context
$(x : A)\ (y : B)$ as $?m\ x\ y$, where $?m$ is a fresh
metavariable. The type of $?m$ is $\uPi (x : A)\ (y : B), C$, where
$C$ is the expected type for the hole at that position. If the
expected type is also unknown at preprocessing time, we create
another fresh metavariable $?m_t : \uPi (x : A)\ (y : B), \Type\ ?u$,
where $?u$ is a fresh universe metavariable. This gives us
$?m : \uPi (x : A)\ (y : B), ?m_t\ x\ y$.  We say a term is
\emph{fully elaborated} if it does not contain metavariables.

We say a term is $\beta$-reducible if it is of the form
$(\lambda x : A, s) t$, and $\iota$-reducible if it is of the form
$\ttt{C.rec}\ \bm{s}\ (\ttt{C.mk}_i\ \bm{r})\ \bm{t}$, where
$\ttt{C.rec}$ is the recursor/eliminator for an inductive datatype
$\ttt{C}$. Here, the sequence $\bm{s}$ represents the parameters,
minor premises and indices, and $(\ttt{C.mk}_i\ \bm{r})$ is the main
premise (where $\ttt{C.mk}_i$ is the $i$-th constructor of $\ttt{C}$).
The function $\ttt{reduce}_{\beta\iota}\ s$ applies head $\beta$ and
$\iota$ reduction to $s$.  We say a term $t$ is \emph{stuck} if
computation cannot occur without instantiating a metavariable $?m$,
where $(?m\ \bm{s})$ is a sub-term of $t$. In that case, we say
$(?m\ \bm{s})$ is the \emph{reason} for $t$ being stuck.  More
formally, a term is stuck when the head symbol is a metavariable
(i.e., it is of the form $?m\ \bm{s}$), or it is a recursor
application where the main premise is stuck.  We say the first case is
a \emph{stuck application}, and the second a \emph{stuck recursor}.

During the preprocessing step, \emph{unification} and \emph{choice}
constraints are generated. Unification constraints are used to enforce
typing constraints, and \emph{choice} constraints are for overloading,
coercion resolution, and triggering the type class mechanism.

A unification constraint $t \ueq s$ is annotated with a
\emph{justification}, which represent the facts and assumptions that
gave rise to the constraint. Justifications are used to assist the generation
of error messages when a term fails to be elaborated, and to implement
\emph{non-chronological backtracking}~\cite{rossi2006handbook}.
Non-chronological backtracking allows exploring the
(possibly infinite) tree of potential solutions
more efficiently, by eliminating branches which we know cannot
possibly contain an actual solution.

There are three kinds of justifications: \emph{asserted},
\emph{assumption} and \emph{join}.  An asserted justification is used
to annotate constraints generated during the preprocessing
phase. Whenever the solver has to perform a choice (also known as a
\emph{case split}), it annotates each choice with a fresh assumption.
A join justification $\join{j_1}{j_2}$ represents the ``union'' of the
justifications $j_1$ and $j_2$. We use $\ucnstr{t}{s}{j}$ to denote
the unification constraint justified by $j$.  A \emph{substitution} is
a finite collection of \emph{assignments} from metavariables to pairs
$\langle t, j\rangle$, written $?m \mapsto \langle t, j\rangle$, where
$t$ is a closed term and $j$ is a justification for the assignment.
Assignments are generated when solving unification constraints.  For
example, the constraint $\ucnstr{?m}{t}{j}$ is solved by adding the
assignment $?m \mapsto \langle t, j\rangle$.  Whenever we apply a
substitution we use a join justification to track its effect. For
example, the result of applying the assignment $?m \mapsto \langle t,
j_m\rangle$ over the constraint $\ucnstr{r}{s}{j}$ is the new
constraint $\ucnstr{r[?m := t]}{s[?m := t]}{\join{j}{j_m}}$.  We also
use $\join{\ucnstr{s}{t}{j_1}}{j_2}$ to denote the constraint
$\ucnstr{s}{t}{\join{j_1}{j_2}}$. Moreover, if $a$ is a list of
constraints $[c_1, \ldots, c_n]$, $\join{a}{j}$ is
$[\join{c_1}{j}, \ldots, \join{c_n}{j}]$.

A \emph{choice constraint} is of the form $\jchoice{?m\ \bm{\ell} :
  t}{f}{j}$, where:
\begin{itemize}
\item $?m$ is a metavariable,
\item $\bm{\ell}$ are free variables representing the context where
  $?m$ was created,
\item $t$ is the type of $?m\ \bm{\ell}$, and
\item $f$ is a procedure that, given the term $?m\ \bm{\ell}$, its
  type $t$, and a substitution, produces a (possibly unbounded) stream
  of constraints representing possible ways of synthesizing $?m$, and
  a justification $j$.
\end{itemize}
Note that each alternative is itself a list of constraints, and is not
necessarily just a \emph{single} unification constraint.

Whereas some constraints should be solved eagerly, other constraints
should be solved only when there is sufficient information to process
them in a reliable way. To that end, a choice constraint
$\bchoice{?m\ \bm{\ell} : t}{f}$ may be marked as \emph{ondemand}.
When the flag \emph{ondemand} is set, the constraint solver will try
to invoke function $f$ only after all metavariables in $t$ have been
instantiated.  We say a \emph{ondemand} choice constraint is
\textbf{ready} when $t$ does not contain metavariables, and
\textbf{postponed} otherwise.  We will later describe how this
feature is used to implement the type class mechanism and coercions.
If a choice constraint is not marked as \emph{ondemand}, we say it is
a \textbf{regular} choice constraint. We use \textbf{regular} choice
constraints to specify overloaded symbols. The result of applying the
assignment $?m \mapsto \langle s, j_m\rangle$ over the choice
constraint $\jchoice{?n\ \bm{\ell} : t}{f}{j}$ is the new constraint
$\jchoice{?n\ \bm{\ell} : t[?m := s]}{f}{\join{j}{j_m}}$. We also use
the notation $\join{c}{j}$ when $c$ is a choice constraint.

\subsection{Support functions}
\label{subsection:support}

In this section, we describe some auxiliary functions that are used
throughout the elaboration algorithm.

The function $\ttt{typeof}\ r$ returns the inferred type of a term
$r$, where $r$ may contain metavariables. Specifically, it returns a
pair $\langle t,\ S\rangle$ where $t$ is the type of $r$ and $S$ is a
set of constraints on the metavariables. If $r$ does not contain
metavariables, then $S$ is empty.

% Leo: Do we need the following explanation?
Given $(\ell_1 : A_1)\ldots(\ell_n : A_n)$, the operation
$\ttt{abstract}_\lambda\ [\ell_1 \ldots \ell_n]\ t$ returns
\[
\lambda (x_1 : A_1)\ldots(x_n : A_n[\ell_1 := x_1,\ldots, \ell_{n-1}
  := x_{n-1}]),\ t[\ell_1 := x_1, \ldots, \ell_n := x_n]
\]
We also have $\ttt{abstract}_\uPi$, the equivalent operation for
$\uPi$-abstraction.

The function $\ttt{unfold}\ (f\ t_1\ldots t_n)$ applies a
$\delta$-reduction, i.e.~it unfolds the definition of constant $f$.
In practice, however, it is not feasible to apply $\delta$-reduction
to all constants in a constraint solving problem. To cope with this
performance issue, we allow the user to annotate definitions with the
hints \emph{irreducible}, \emph{semireducible} or \emph{reducible}, as
described in Section~\ref{subsection:computational}. Recall that an
irreducible definition is never unfolded by the constraint solver,
while a semireducible or reducible definition may be unfolded or not
depending on the constraint being solved. Roughly, a semireducible
definition is unfolded only if the decision to unfold is ``simple,''
which is to say, if the unfolding does not require the procedure to
consider an extra case split. When a decision is not simple, the
unfolding produces at least one extra case, and consequently increases
the search space. When no annotation is provided, the system assumes
the definition is \emph{semireducible}. Note that when the kernel type
checks fully elaborated definitions, these annotations are ignored;
they are only relevant during the elaboration process.

The function $\ttt{whnf}\ r$ returns a pair $\langle w,\ S\rangle$
where $w$ is a term convertible to $r$ that is in weak head normal
form (\emph{whnf}) or is stuck, and $S$ is a set of unification
constraints.  (In this paper, we can assume the set $S$ returned by
$\ttt{whnf}$ is always empty. In our implementation, constraints in
$S$ arise because the elimination rule for equality has extra
computational reductions when proof irrelevance is enabled,
corresponding to Streicher's axiom K \cite{streicher:93}.
Specifically, given \ttt{H₁~:~a = a}, the term \ttt{eq.rec A a C a H₂
  H₁} reduces to \ttt{H₂} even when \ttt{H₁} is not the constructor
\ttt{eq.refl A a}, where \ttt{eq.rec} is the recursor for the Leibniz
equality. To that end, the elaborator has to infer the type of
\ttt{H₁}, which can give rise to unification constraints.) The
function $\ttt{whnf}$ does not unfold irreducible definitions, and
during type class resolution, it also does not unfold semireducible
ones.

The procedure $\ttt{error}\ j$ throws an exception tagged with a
justification $j$.

Finally, the function $\ttt{ensurefun}\ s\ j$ ensures that $s$ has a
function type. Specifically, it infers the type $t$ of $s$ (using
\ttt{typeof}) and then reduces $t$ to $t'$ in weak head normal form.
If $t'$ is a $\uPi$-term, then it returns $t'$ and any new unification
constraints.  If $t'$ is not a $\uPi$-term and is not stuck, then it
generates an error with justification $j$.  Otherwise, if $?m\ \bm{s}$
is the reason that $t'$ is stuck, where
$?m : (\uPi \bm{x} : \bm{A}, B)$, we create two fresh metavariables:
$?m_1 : (\uPi \bm{x} : \bm{A},\ \Type\ ?u_1)$ and
$?m_2 : (\uPi (\bm{x} : \bm{A})\,(y :\, ?m_1\ \bm{x}),\ \Type\ ?u_2)$,
and the new constraint
\[
\ucnstr{t}{(\uPi x :\, ?m_1\ \bm{s},\ ?m_2\ \bm{s}\ x)}{j}.
\]
This ensures that $s$ has a function type, and defers the problem of
figuring out what that type is.

Recall that we use the \emph{local nameless} approach for representing
terms, in which free variables have a unique identifier and a type,
while bound variables are represented by a de Bruijn index. We say a
term $t$ has a \emph{dangling} bound variable if there is a bound
variable in $t$ that is not in the scope of any
$\lam$/$\uPi$-expression binding it. For example, the
$\lam$-expression $\lam x : \ttt{Type}, f\ x\ (g\ x)$ has no dangling
bound variables, but $x$ is a dangling bound variable in
$f\ x\ (g\ x)$.  In the locally nameless approach, all major
operations (such as type inference, normalization, and unification)
assume there are no dangling bound variables, that is, no bound
variables that ``point out of the scope.'' This invariant is enforced
by replacing bound variables with fresh free variables whenever we
visit the body of a $\lam$/$\uPi$-expression. In the lambda expression
above, we would replace $x$ in $f\ x\ (g\ x)$ with the free variable
$\ell : Type$. This operation is called $\ttt{instantiate}$ in
\cite{IamNotANumber}.  The operation $\ttt{abstract}\ \ell\ t$ is the
inverse; it replaces the free variable $\ell$ in $t$ with the bound
variable with de Bruijn index $0$.  These two operations are
essentially the only ones that have to deal with de Bruijn indices.

Although the locally nameless approach greatly simplifies the
implementation effort, there is a performance penalty.  Given a term
$t$ of size $n$ with $m$ binders, it takes $O(n m)$ time to visit $t$
while making sure there are no dangling bound variables.
In~\cite{IamNotANumber}, the authors suggest that this cost can be
minimized by generalizing $\ttt{abstract}$ and $\ttt{instantiate}$ to
process sequences of free and bound variables. This optimization is
particularly effective when visiting terms containing several
consecutive binders, such as $\lam x_1 : A_1, \lam x_2 : A_2, \ldots,
\lam x_n : A_n, t$.  Nonetheless, we observed that these two
operations were still a performance bottleneck for several files in
the Lean standard library.  We have addressed this problem using a
very simple complementary optimization.  For each term $t$, we store a
\emph{bound} $B$ such that all de Bruijn indices occurring in $t$ are
in the range $[0, B)$.  This bound can easily be computed when we
  create new terms: the bound for the de Bruijn variable with index
  $n$ is $n+1$, and given terms $t$ and $s$ with bounds $B_t$ and
  $B_s$ respectively, the bound for the application $(t\ s)$ is
  $\ttt{max}(B_t, B_s)$, and the bound for $(\lam x : t, s)$ is
  $\ttt{max}(B_t, B_s - 1)$. We use the bound $B$ to optimize the
  $\ttt{instantiate}$ operation. The idea is simple: $B$ enables us to
  decide quickly whether any subterm contains a bound variable being
  instantiated or not. If it does not, then our $\ttt{instantiate}$
  procedure does not even visit the subterm.  Similarly, for each term
  $t$, we store a bit that is set to ``true'' if and only if $t$
  contains a free variable.  We use this bit to optimize the
  $\ttt{abstract}$ operation, since it enables us to decide quickly
  whether a subterm contains a free variable.

These optimizations are crucial to our implementation. The Lean
standard library currently contains 172 files, and 41,700 lines of
Lean code.  With the optimizations, the whole library can be compiled
in 71.06 seconds using an Intel Core i7 3.6Ghz processor with 32Gb of
memory. Without the optimizations, it takes 2,189.97 seconds to
compile the same set of files.

\subsection{The constraint simplification procedure}
\label{subsection:simplification}

The preprocessing step and the constraint-solving procedure rely on a
constraint-simplification procedure, which we now describe. The idea
of the \ttt{simp} procedure is to decompose all the constraints that
can ``straightforwardly'' be decomposed to simpler ones, and to detect
quickly any constraints that simply cannot be solved. Thus, given a
unification constraint, the \ttt{simp} procedure produces a set of
(potentially) simpler unification constraints or throws an error.
Moreover, if the input constraint does not contain metavariables, then
the result is the empty set $\{\}$ or an error.

In the pseudocode below, $s$ and $t$ denote arbitrary terms, $\ell$ is
a free variable, and $f$ and $g$ are constants. The procedure
$\ttt{mklocal}\ A$ creates a fresh free variable with type $A$.  To
simplify the presentation, we assume there is a global unique name
generator.  The function $\ttt{depth}\ f$ returns the \emph{definition
  depth} of the constant $f$, which is equal to 0 if $f$ is not a
definition, and $1 + \ttt{max} \{ \ttt{depth}\ g\ |\ g\ \text{appears
  in the definition of}\ f\}$ otherwise.  To save space, we do not
list symmetric cases; for example, we present a case for
$\ucnstr{s}{(\lambda x : B, t)}{j}$ but not $\ucnstr{(\lambda x : B,
  t)}{s}{j}$.

\begin{tabbing}
% Fake line for setting tabs
\ttt{si} \= \ttt{mp} \= \ttt{lify} $\ucnstr{s}{t}{j}$ \bwhen\ $s$ is reducible \= = \kill
\ttt{simp} $\ucnstr{t}{t}{j}$ \> \> \> = \{\} \\
\ttt{simp} $\ucnstr{s}{t}{j}$ \bwhen\ $s$ is $\beta/\iota$-reducible \> \> \> =
  \ttt{simp} $\ucnstr{\ttt{reduce}_{\beta\iota}\ s}{t}{j}$ \\
\ttt{simp} $\ucnstr{\ell\ s_1 \ldots s_n}{\ell\ t_1 \ldots t_n}{j}$ \> \> \> =
  $\bigcup_{i=1}^n \ttt{simp}\ \ucnstr{s_i}{t_i}{j}$ \\
\ttt{simp} $\ucnstr{f\ s_1 \ldots s_n}{f\ t_1 \ldots t_n}{j}$ \> \> \> = \\
  % Do we need the following extra
  \> \bif\ $s_1 \ldots s_n$ and $t_1 \ldots t_n$ do not contain metavariables \bthen \\
  \> \> \ttt{simp} $(\ucnstr{\ttt{unfold}\ (f\ s_1 \ldots s_n)}{\ttt{unfold}\ (f\ t_1 \ldots t_n)}{j})$ \\
  \> \belseif\ $f$ is not \emph{reducible} \bthen\ $\bigcup_{i=1}^n \ttt{simp}\ \ucnstr{s_i}{t_i}{j}$ \\
  \> \belse\ $\{\ucnstr{f\ s_1 \ldots s_n}{f\ t_1 \ldots t_n}{j}\}$ \\
\ttt{simp} $\ucnstr{f\ \bm{s}}{g\ \bm{t}}{j}$ \> \> \> = \\
  \> \bif\ \ttt{depth} $f > \mbox{\ttt{depth}}\ g$ and $f$ is not \emph{irreducible} \bthen \\
  \> \> \ttt{simp} $(\ucnstr{\ttt{unfold}\ (f\ \bm{s})}{g\ \bm{t}}{j})$ \\
  \> \belseif\ \ttt{depth} $f < \mbox{\ttt{depth}}\ g$ and $g$ is not \emph{irreducible} \bthen \\
  \> \> \ttt{simp} $(\ucnstr{f\ \bm{s}}{\ttt{unfold}\ (g\ \bm{t})}{j})$ \\
  \> \belseif\ \ttt{depth} $f$ = \ttt{depth} $g$ and $f$ and $g$ are not \emph{irreducible} \bthen \\
  \> \> \ttt{simp} $(\ucnstr{\ttt{unfold}\ (f\ \bm{s})}{\ttt{unfold}\ (g\ \bm{t})}{j})$ \\
%  \> \belse\ \ttt{error} $j$ \\
\ttt{simp} $\ucnstr{(\lambda x : A, s)}{(\lambda y : B, t)}{j}$ \> \> \> = \\
  \> \blet\ $\ell$ = \ttt{mklocal} $A$
     \bin\ \ttt{simp} $\ucnstr{A}{B}{j}$ $\cup$ \ttt{simp} $\ucnstr{s[x := \ell]}{t[y := \ell]}{j}$ \\
\ttt{simp} $\ucnstr{(\uPi x : A, s)}{(\uPi y : B, t)}{j}$ \> \> \> = \\
  \> \blet\ $\ell$ = \ttt{mklocal} $A$
     \bin\ \ttt{simp} $\ucnstr{A}{B}{j}$ $\cup$ \ttt{simp} $\ucnstr{s[x := \ell]}{t[y := \ell]}{j}$ \\
\ttt{simp} $\ucnstr{s}{(\lambda x : B, t)}{j}$ \> \> \> = \\
  \> \blet\ $\langle (\uPi x : A, C), S \rangle$ = \ttt{ensurefun} $s$ $j$ \bin\ \\
  \> \> \ttt{simp} $\ucnstr{(\lambda x : A, s\ x)}{(\lambda x : B, t)}{j} \cup S$ \\
\ttt{simp} $\ucnstr{s}{t}{j}$    \> \> \> = \\
  \> \bif\ $s$ or $t$ is \emph{stuck} \bthen\ $\{\ucnstr{s}{t}{j}\}$ \belse\ \ttt{error} $j$
\end{tabbing}

In the actual implementation, we also use a heuristic optimization for
the case \ttt{simp} $\ucnstr{f\ s_1 \ldots s_n}{f\ t_1 \ldots
  t_n}{j}$, where $s_1 \ldots s_n$ and $t_1 \ldots t_n$ do not contain
metavariables, and $f$ is not a projection. % TODO: define projection
In this case, we first try \ttt{simp} $\ucnstr{s_1}{t_1}{j}$ \ldots
\ttt{simp} $\ucnstr{s_n}{t_n}{j}$, and if no error is thrown, we
return $\{\}$.

Each unification constraint returned by \ttt{simp} is in one of the
following categories:
\begin{itemize}
\item \textbf{delta}: $\ucnstr{f\ \bm{s}}{f\ \bm{t}}{j}$.
Note that, based on the definition of \ttt{simp}, $f$ must be a reducible definition.
\item \textbf{pattern}: $\ucnstr{?m\ \ell_1 \ldots \ell_n}{t}{j}$,
where $\ell_1, \ldots, \ell_n$ are pairwise distinct free variables, $t$ only
contains free variables in $\{\ell_1, \ldots, \ell_n\}$, and $?m$ does not occur in $t$.
\item \textbf{quasi-pattern}: $\ucnstr{?m\ \ell_1 \ldots \ell_n}{t}{j}$, where
all $\ell_1, \ldots, \ell_n$ are free variables, but are not pairwise distinct.
\item \textbf{flex-rigid}: $\ucnstr{?m\ s_1 \ldots s_n}{t}{j}$, where
at least one of $s_1, \ldots, s_n$ is not a free variable.
\item \textbf{flex-flex}:  $\ucnstr{?m_1\ \bm{s}}{\,?m_2\ \bm{t}}{j}$.
\item \textbf{recursor}:  $\ucnstr{t}{s}{j}$, where $t$ or $s$ is a
  stuck recursor.
\end{itemize}
In the literature, \textbf{pattern}, \textbf{quasi-pattern}, and
\textbf{flex-rigid} are simply called flex-rigid constraints, and the
category \textbf{pattern} corresponds to Miller
patterns~\cite{miller:nadathur:12}. Note that flex-flex constraints
are badly underconstrained, and we typically expect that other
constraints will do more to limit the interpretation of the
metavariables.

\subsection{Preprocessing}
\label{subsection:preprocessing}

%% The parser produces preterms that would not even type check
%% in the kernel because they correspond to partial constructions.
%% For example, consider the identity function, \ttt{id :
%%   $\uPi$\{A : Type\}, A → A}, from the standard library, the first
%% argument is marked as implicit and does not need to be explicitly
%% provided by users. Suppose the user inputs \ttt{id zero}, this
%% preterm would not type check in the kernel because the first argument of \ttt{id} is expected
%% to be a type.
The preprocessor is a recursive procedure that, given a preterm and a
context, returns a term $t$ (potentially containing metavariables) and
a set of unification and choice constraints. The basic idea is that if
the constraints are solved, their solution should contain an
assignment for all metavariables in $t$.  The preprocessor must carry
a context, a list of free variables, to be able to create fresh
metavariables. This is the only procedure in our implementation that
``carries contexts around.''  The preprocessor only creates
\emph{asserted} justification objects.

Applications $(r\ s)$ are the main source of unification constraints.
After a preterm $p$ in a context $\bm{\ell}$ is converted into the
application $(r\ s)$, the preprocessor uses \ttt{ensurefun} to make
sure that the type of $r$ is of the form $\uPi x : A,B$, and
\ttt{simp} to enforce that the type $C$ of $s$ is convertible to $A$.
If $C$ is not convertible to $A$, the preprocessor checks the database
of available coercions. If there is a coercion $c$ from $C$ to $A$, it
replaces the application $(r\ s)$ with $(r\ (c\ s))$.  If $A$ is
stuck, but there are coercions $\{c_1, \ldots, c_n\}$ from $C$, the
preprocessor creates a fresh metavariable $?m\ :
\ttt{abstract}_\uPi\ \bm{\ell}\ A$, replaces the application with
$(r\ (?m\ \bm{\ell}))$, and creates a \emph{ondemand} choice
constraint $\jchoice{?m\ \bm{\ell} : A}{f}{j}$, where the choice
function $f$ produces one of the following alternatives $s$, $c_1\ s$,
\ldots, $c_n\ s$.  If possible, the solver will only invoke $f$ after
all metavariables in $A$ have been instantiated. In this ideal
situation, $f$ returns at most one solution, and no case-analysis is
needed.
%% However,
%% in some cases, this is not possible, consider the term $\ttt{eq}\ ?A\ n\ m$
%% where $n$ and $m$ have type $\ttt{nat}$ and there is a coercion from $\ttt{nat}$
%% to $\ttt{int}$.
The same process is performed when $C$ is stuck and there are
coercions \emph{to} $A$.  We currently do \emph{not} try to inject
coercions when both $A$ and $C$ are stuck at preprocessing time.

As noted in Section~\ref{subsection:coercions}, \lean{} supports
parametric coercions, and coercions to sorts and function classes. Ad
hoc overloading is also realized using choice constraints. The idea is
the same, but we create a \emph{regular} choice constraint, where the
choice function $f$ produces the different interpretations for the
overloaded symbol.

In a context $\bm{\ell}$, a placeholder ``\ttt{\_}'' is simply replaced by
$?m\ \bm{\ell}$, where $?m$ is a fresh metavariable.

Finally, to handle implicit arguments, when we infer the type $t$ of a
term $r$, if $t$ is of the form $\uPi\{x : A\},B$, then we create a fresh
metavariable $?m : \ttt{abstract}_\uPi\ \bm{\ell}\ A$ and replace $r$
with the application $(r\ (?m\ \bm{\ell}))$. If the implicit argument
is marked with square brackets to indicate it should be synthesized by
the type class mechanism, we also create an \emph{ondemand} choice
constraint $\jchoice{?m\ \bm{\ell} : A}{f}{j}$, where the choice
function $f$ invokes the type class resolution procedure.  This
procedure is essentially a simple $\lambda$-Prolog
interpreter~\cite{miller:nadathur:12}, where the Horn clauses are the
user-declared instances.

\subsection{The constraint solving procedure}
\label{subsection:solving}

Given a set of constraints, our solver returns a failure, or a
substitution $S$ and set of \textbf{flex-flex} constraints of the form
$\ucnstr{?m_1\ \bm{s}}{\,?m_2\ \bm{t}}{j}$ such that neither $?m_1$
nor $?m_2$ are assigned in $S$. In other words, it is required to
solve all the constraints that are presented to it, but it does not
assign metavariables whose solutions are underconstrained.

The solver uses the following data structures:
\begin{itemize}
\item a priority queue $Q$ of constraints,
\item a mapping $U$ of metavariables to constraints,
\item a substitution $S$, and
\item a case split stack $C$.
\end{itemize}
To simplify the presentation, we assume $Q$, $U$, $S$ and $C$ are global variables.

The priorities for the $Q$ are computed using the following total order, $\prec$,
on constraint categories:
\begin{quote}
  \textbf{pattern} $\prec$
  \textbf{ready} $\prec$
  \textbf{regular} $\prec$
  \textbf{delta} $\prec$
  \textbf{quasi-pattern} $\prec$ \\
  \hspace*{0.2in}
  \textbf{flex-rigid} $\prec$
  \textbf{recursor} $\prec$
  \textbf{postponed} $\prec$
  \textbf{flex-flex}
\end{quote}
Recall that \textbf{ready}, \textbf{regular}, and \textbf{postponed}
are all choice constraints. If two constraints are in the same
category, we use the \emph{first-in-first-out} method.

The mapping $U$ works as follows: for each metavariable $?m$, $U[?m]$
is the finite subset of the constraints in $Q$ such that for each $c$
in $U[?m]$, $c$ is a unification constraint stuck because of $?m$, or
$c$ is an \emph{ondemand} choice constraint $\jchoice{?n\ \bm{\ell} :
  t}{f}{j}$ and $?m$ occurs in $t$.  The set $U[?m]$ contains the set
of constraints that need to be (re-)visited whenever $?m$ is
assigned. We remark that a unification constraint in $U[?m]$ may
become simpler after we replace $?m$ with its assignment. Similarly,
an \emph{ondemand} choice constraint $\jchoice{?m\ \bm{\ell} :
  t}{f}{j}$ in $U[?m]$ is ready to be processed when all metavariables
in $t$ have been assigned.

Given a set of constraints $s$, for each constraint $c$ in $s$, the
procedure $\ttt{visit}\ s$ simply invokes $\ttt{visiteq}\ c$ if $c$ is
a unification constraint, and $\ttt{visitchoice}\ c$ otherwise.  The
procedure $\ttt{visiteq}\ \ucnstr{r}{s}{j}$ is defined as follows:
\begin{tabbing}
== \= == \= ===== \kill
\> \bif\ $r$ or $s$ is stuck by some $?m$ and $?m \mapsto \langle t, j_m \rangle$ in $S$ \bthen \\
   \> \> $\ttt{visit}\ (\ttt{simp}\ \ucnstr{r[?m := t]}{s[?m := t]}{\join{j}{j_m}})$ \\
\> \belseif\ the constraint is a \textbf{pattern} $\ucnstr{?m\ \bm{\ell}}{t}{j}$ \bthen \\
   \>  \> add the assignment $?m \mapsto \langle (\ttt{abstract}_\lambda\ \bm{\ell}\ t), j \rangle$ to $S$ \\
   \>  \> for each \ttt{c} in $U[?m]$, $\ttt{visit}\ (c)$ \\
\> \belse\ update $U$, and insert constraint into $Q$
\end{tabbing}
The procedure $\ttt{visitchoice}\ \jchoice{?n\ \bm{\ell} : t}{f}{j}$ just substitutes any assigned metavariable
$?m$ occurring in $t$, updates $U$, and inserts the constraint into $Q$. Note that, we never insert
\textbf{pattern} constraints into $Q$.

To implement a backtracking search, we need a mechanism for restoring
the state of the solver during a backtrack operation. We use a very
simple approach where $Q$, $U$, and $S$ are implemented using pure data
structures (red-black trees) that provide a constant time copy
operation.  Whenever we need to create a case split, we simply create
copies of $Q$, $U$ and $S$.  An alternative approach is to use a
\emph{trail stack}~\cite{rossi2006handbook} which stores operations
that ``undo'' the destructive updates performed during the search. We
have determined that our simpler approach for implementing
backtracking is not a bottleneck in our implementation.

When solving a non-\textbf{pattern} constraint $c$, the solver creates a
case split, and stores it on the stack $C$. Each case split is a tuple
of the form $\langle Q_c, U_c, S_c, j_a, j_c, z \rangle$, where
\begin{itemize}
\item $Q_c$, $U_c$ and $S_c$ store the state of the solver when the
  case split was created,
\item $j_a$ is a fresh assumption justification used to track the case
  split,
\item $j_c$ is the justification for $c$, and
\item $z$ is a lazy list containing the remaining alternatives, where
  each alternative is a list of constraints.
\end{itemize}
We use $\ttt{pull}\ z$ to denote the operation that destructively
extracts the head of the lazy list $z$ and returns it, or returns
$\ttt{none}$ when $z$ is empty.  The solver catches any
$\ttt{error}\ j$ thrown by the \ttt{simp} procedure, and uses the
error resolution procedure $\ttt{resolve}\ j$ defined as follows:
\begin{tabbing}
== \= = \= = \= = \= ============= \= \kill
\> \bwhile\ $C$ is not empty \\
\> \> \blet\ $\langle Q_c, U_c, S_c, j_a, j_c, z \rangle$ = \ttt{top}\ $C$ \bin \\
\> \> \bif\ $j$ depends on $j_a$ \bthen\ \\
\> \> \> restore state $Q$ := $Q_c$, $U$ := $U_c$, $S$ := $S_c$ \\
\> \> \> \bif\ $\ttt{pull}\ z = \ttt{some}\ a$ \bthen\ $\ttt{visit}\ (\join{\join{a}{j_c}}{j})$ and \breturn\ \ttt{pop} $C$ \\
\> \textbf{failed} to solve constraints since $C$ is empty
\end{tabbing}
In the procedure above, $\ttt{visit}\ (\join{\join{a}{j_c}}{j})$ may throw another $\ttt{error}\ j'$.
If this happens it recursively invokes $\ttt{resolve}\ j'$.

\subsection{Processing constraints}
\label{subsection:processing}

At the very core of the algorithm is the procedure for processing the
constraints in the the queue $Q$, which we now describe. We use an
auxiliary procedure $\ttt{process}\ z\ j$, where $z$ is a lazy list of
alternatives, and $j$ is a justification. If $z$ is empty, it just
invokes $\ttt{resolve}\ j$.  Otherwise, it pulls the head $a$ of $z$,
creates a fresh assumption justification $j_a$, pushes the new case
split $\langle Q, U, S, j_a, j_c, z \rangle$ on the stack $C$, and
invokes $\ttt{visit}\ (\join{\join{a}{j_a}}{j})$.

For choice constraints $\jchoice{?m\ \bm{\ell} : t}{f}{j}$, whether
they are \textbf{ready}, \textbf{regular} or \textbf{postponed}, we
just invoke $\ttt{process}\ (f\ (?m\ \bm{\ell})\ t\ S)\ j$.

For \textbf{delta} constraints $\ucnstr{f\ s_1 \ldots s_n}{f\ t_1\ldots
  t_n}{j}$, we try two alternatives. In the first one, we assume $f$
is opaque, and try to avoid the potentially expensive
$\delta$-reduction step by using $a_1 = \bigcup_{i=1}^n
\ttt{simp}\ \ucnstr{s_i}{t_i}{j}$. If it fails, as our next
alternative, we unfold $f$ and try $a_2 = \ttt{simp}
(\ucnstr{\ttt{unfold}\ (f\ s_1 \ldots s_n)}{\ttt{unfold}\ (f\ t_1
  \ldots t_n)}{j})$.  We use the operation \ttt{tolazy} to convert the
list $[a_1, a_2]$ into a lazy list, and process the delta constraint
using $\ttt{process}\ (\ttt{tolazy}\ [a_1, a_2])\ j$. This case split
is a heuristic optimization and is not necessary for completeness.

The two constraint categories \textbf{quasi-pattern} and
\textbf{flex-rigid} are handled in the same way; we use different
categories only to ensure that easier constraints occur first in the
priority queue. We undertake an incomplete search for solutions to
these constraints using a variation of the flex-rigid case of Huet's
unification algorithm~\cite{huet75}.  Given a flex-rigid constraint
$\ucnstr{?m\ s_1\ldots\ s_p}{t}{j}$, the main idea behind Huet's
algorithm is the observation that $t$ must be a term of the form
$f\ r_1\ldots r_n$, where $f$ is a free variable or a constant. The
next idea is the observation that any solution for $?m$ is convertible
to one in eta-long normal form, which allows us to consider only
solutions for $?m$ that are of the form
\begin{equation}
\tag{*}
\lambda x_1 \ldots x_n, h\ (?m_1\ x_1 \ldots x_n) \ldots (?m_p\ x_1 \ldots x_n)
\end{equation}
where $?m_i$ are fresh metavariables, and $h$ is a constant or one of
the bound variables $x_1 \ldots x_n$.

In Huet's algorithm, only opaque constants are considered, so if $h$
is a constant different from $f$ of the rigid term $t$, the solution
would lead to an unsolvable constraint. Therefore, we say that Huet's
procedure has two kinds of case splits: \emph{imitation} (when $h$ is
the constant $f$ of the rigid term), and \emph{projection} (when $h$
is one of the bound variables $x_1 \ldots x_n$). However, there are
two complications in our setting. First, we do not eagerly unfold
$f\ r_1\ldots r_n$ when $f$ is a constant. For example, assume that
$\ttt{sub}\ a\ b$ (subtraction for integers) is defined as
$\ttt{add}\ a\ (\ttt{uminus}\ b)$.  Then
$\ucnstr{?m\ (\ttt{uminus}\ a)}{\ttt{sub}\ b\ a}{j}$ has a solution
$?m = \lambda x, \ttt{add}\ b\ x$, but we would miss it if we did not
unfold $\ttt{sub}$ before trying to imitate.  Second, we have
recursors in our language, and even if $f$ is an opaque constant, it
is not the only constant that can be used for $h$. For example, given
the constraints $\ucnstr{?m\ \ttt{zero}}{\ttt{true}}{j},
\ucnstr{?m\ (\ttt{succ}\ \ttt{zero})}{\ttt{false}}{j}$, a possible
solution is $?m = \lambda x,\ \ttt{nat.rec}\ (\lambda n,
\ttt{bool})\ \ttt{true}\ (\lambda n\ r, \ttt{false})\ x$, where
\ttt{nat.rec} is the recursor for the type \ttt{nat} (of the natural
numbers).  We cope with the first problem using an approach similar to
the one used for \textbf{delta}-constraints when $f$ is a
\emph{reducible} constant. The idea is to have two imitation steps,
one where $f$ is not unfolded, and another one where the term
$f\ r_1\ldots r_n$ is put into weak head normal form before performing
the imitation.  In our implementation, it is currently infeasible to
consider the extra imitation step (after \ttt{whnf}) for all
constants.  Even using non-chronological backtracking, the search
space becomes too big. The main problem is that the system may spend a
huge amount of time traversing the whole search space when the user
provides an incorrect partial construction. As to the second issue, we
currently simply ignore this possiblity, since the search space would
become too big if we considered recursors for $h$. Moreover, if $h$ is
a recursor, the constraint obtained after replacing $?m$ would be a
stuck recursor.

As in most higher-order unification procedures, we try first the
projection case splits because they generate more general
solutions. We remark that the number of case splits can usually
be greatly reduced for \textbf{quasi-pattern}s, which is the case the
arises most commonly in practice. In this case, if $f$ is a constant
(not marked as reducible), then we do not need to consider any
projections. Any projection would fail immediately: if we take $h$ to
be $\ell_i$ and substitute (*) for $?m$ in the original constraint, we
obtain an unsolvable constraint
$\ucnstr{\ell_i\ (?m_1\ \bm{\ell})\ldots
  (?m_p\ \bm{\ell})}{f\ r_1\ldots r_n}{j'}$. Finally, if $f$ is a free
variable $\ell$, then we only need to consider the projection where
$h$ is $x_i$ if $\ell_i = \ell$.

For \textbf{flex-rigid} constraints $\ucnstr{?m\ s_1 \ldots
  s_n}{t}{j}$, we only consider the case $h$ is $x_i$ when $s_i$ is a
free variable $\ell$, or $s_i$ is convertible to $t$. In the second
case, where $s_i$ is convertible to $t$, we simply assign $\lambda x_1
\ldots x_n, x_i$ to $?m$.  This is a heuristic for reducing the size
of the state, and minimizing the number of instances where the
procedure exhibits nonterminating behavior.  We note that in the
second-order case, the solver does not miss solutions by using this
heuristic.  Finally, our solver has a threshold on the number of steps
that can be performed.

We also use an approximate solution for \textbf{recursor} constraints
$\ucnstr{t}{s}{j}$. If the head of $t$ and $s$ is the same recursor
\ttt{C.rec}, then we try to solve the constraint by treating
\ttt{C.rec} as a regular opaque constant which has no computational
behavior associated with it.  If $t$ or $s$ is of the form
$?m\ \bm{r}$, then we treat it as a flex-rigid constraint.  In a
previous implementation of our algorithm, when the recursor
\ttt{C.rec} was stuck because of a term $?m\ \bm{r}$, we tried to
perform a case split for each constructor $\ttt{C.mk}_i$ of $\ttt{C}$,
assigning $?m$ to terms of the form $\lambda \bm{x},
\ttt{C.mk}_i\ (?m_1\ \bm{x}) \ldots (?m_n\ \bm{x})$. However, this
provides only a minor improvement on the usability of the system: only
three theorems in our library broke after we removed this feature, and
all of them could be easily fixed by providing implicit arguments
explicitly.

\section{Related work and conclusions}
\label{section:conclusions}

The elaboration algorithm we have described above has been developed
and tuned in conjunction with the development of \lean{}'s standard
and homotopy type theory libraries. Although these libraries are still
under development, they provide ample evidence that the approach we
describe here is effective in practice. At the time of writing, the
standard library consists of about 42k lines of code, with core
datatypes including products, lists, sets, multisets (bags), tuples,
subtypes, and vectors; core number systems, namely, the natural
numbers, integers, rationals, reals, and complex numbers; algebraic
structures, including orders, (ordered) groups, (ordered) rings,
(ordered) fields, and so on; elementary finite group theory, through
Sylow's theorem; elementary number theory, such as the unique
factorization theorem; the beginnings of analysis, including the
completeness of the reals and elementary properties of limits. The
homotopy type theory library consists of more than 25k lines of code,
including most of the first seven chapters of the Homotopy Type Theory
book \cite{hottbook}, and a substnatial development of category
theory. Specifically, it includes core datatypes and constructions,
such as paths, fibrations, equivalences, and pathovers; higher
inductive types, such as the circle, sphere, torus, quotients,
pushouts, suspensions; the calculation of the homotopy group of the
circle; and category theory through the Yoneda lemma. \lean{} also
supported a substantial development in nonabelian algebraic topology
\cite{von:raumer:15}, carried out by Jakob von Raumer in the homotopy
type theory framework.

%% Historically, more effort has been put into understanding the core
%% type theory, both from a theoretical point of view and for
%% pragmatic implementations then for the elaboration from user input
%% into the core theory. However, higher-order unification, an
%% integral part of the process, has been a subject of extensive
%% study. Let us simply note, in addition to Huet's foundational
%% work~\cite{huet75}, Conal Elliot's PhD
%% dissertation~\cite{Elliot90:Dissertation}, which extends Huet's
%% algorithm to a dependently typed theory with
%% $\Sigma$-types. Higher-order unification is undecidable and lacks
%% the property of \emph{most general unifier}, even if we allow a
%% finite set of most general unifiers instead of a single
%% one. However, Elliot remarks that it is possible to describe
%% reasonable algorithms that enumerate the set of solutions to a
%% higher-order unification problem. This algorithm will not terminate
%% in general, but may perform well in practice. we take this approach
%% in the present work. Our use of justifications to elide entire
%% sections of the search tree is, as far as we can tell, entirely
%% novel in the realm of higher-order unification. It is directly
%% inspired by the method of \emph{non-chronological
%% backtracking}~\cite{rossi2006handbook} in modern constraint solving
%% algorithms.

We attempt to put our work in the context of recent work on
elaboration in dependent type theories. Abel and Pientka present an
extension of Miller-style pattern unification~\cite{abelPientka:tlca11}
which can handle a larger class of problems (in addition to
$\Sigma$-types) by a method they call \emph{pruning}, which,
intuitively, removes arguments to metavariables which fall outside of
the Miller pattern fragment, allowing for more solutions to be
found. They also give a bi-directional inference system for a
dependently typed $\lam$-calculus, which together with the unification
algorithm yields an outline for a practical implementation. They show
the soundness of the unification algorithm with respect to this type
system. They do not, however, treat the case of defined constants,
with or without recursion.

Building upon this is recent work by Ziliani and Sozeau~\cite{ZilianiSozeauUnif}
that describes a unification algorithm for the Coq theorem prover
which features defined constants and recursively defined
functions. They attempt to describe the practicalities of such an
algorithm for a realistic dependently typed language, outlining the
heuristics and efficiency compromises inherent in this task. In that
respect, their motivations are very similar to ours.

In addition to Abel and Pientka's pruning, Ziliani and Sozeau add a
more aggressive form of dependency erasure for metavariables, in an
attempt to solve more unification problems at the cost of uniqueness
of solutions. One example is the problem $\{?t\ \mathrm{true}\ueq
\mathrm{nat},\ ?t\ \mathrm{false}\ueq \mathrm{nat}\}$.  This problem
is solved in their framework by dropping the dependency of $?t$ on its
argument, and returning the constraint $?t'\ueq \mathrm{nat}$ which
gives the solution $?t\mapsto \lx,\mathrm{nat}$.  They also add a
resolution rule called \emph{first order approximation}, in which for
example the constraint ${?f\ ?y}\ueq{S\ 0}$ is solved with the
assignment $?f\mapsto S, ?y\mapsto 0$

Since we have no qualms about allowing multiple solutions and
backtracking search our algorithms can handle both of these problems
easily, in the first case by a special case of \emph{projection}, and
in the second by an \emph{imitation} step.  Our approach to free
variables in metavariables is simple: there are none. In contrast,
Ziliani and Sozeau carry around a suspended substitution with every
metavariable, that needs to be managed in each resolution step.  The
heuristics outlined in their paper for unfolding constants are similar
to ours: constants are unfolded only after an attempt has been made to
apply type-class resolution, and constants are unfolded to a pattern
match or fixpoint only in last resort. More study is needed to examine
the trade-offs of these various choices.  Finally, their system does
not allow postponement of constraints, relying on pruning and
dependency erasure to treat most cases up-front. They argue that great
efficiency gains are obtained in this manner. Again, more study is
required to assess the trade-offs of this approach.

Various algebraic developments in Coq make use of type classes
\cite{sozeau:oury:08,spitters:van:der:weegen:11,gross:et:al:14} and
\emph{canonical structures}
\cite{saibi:97,mathstruct,mahboubi:tassi:13}; see also
\cite{asperti:et:al:09} for the use of \emph{unification hints} in
Matita. Many of the features we have described are also implemented in
systems based on simple type theory. For example, Isabelle uses
axiomatic typeclasses~\cite{wenzel2005using} and parameterized
contexts (locales)~\cite{locales} to deal with algebraic
structures. It also has mechanisms to insert coercions
\cite{traytel:et:al:11}. The reliance on simple type theory, however,
makes the elaboration problem quite different from ours. For example,
an algebraic structure that depends on a parameter, such as the
integers modulo $m$, cannot be represented as a type, and so cannot be
an instance of an axiomatic type class. In contrast to Lean, Isabelle
uses different languages to construct expressions and assertions,
build proofs, and express relationships between structures.

In a different vein, recent work by Brady on the dependently typed
language Idris~\cite{idris} describes the elaboration process by
analogy with theorem proving (and in the context of pure functional
programming). Our work is in stark contrast with his, as our tactic
language is completely disjoint from the methods with which we specify
the constraint resolution for the unification problems. In \lean{},
the problems are quite different: in unification, metavariables can
be very non-local, appearing in disparate contexts and the solutions
can be an infinite stream rather than a simple finite case split.

%% Furthermore, metavariables may appear and be computationally relevant
%% to the term under elaboration, whereas the shape of proofs is
%% typically irrelevant, both in size and in computational content
%% (barring program extraction from proofs).  At the moment we do not
%% know whether there is a satisfactory unifying language for the
%% synthesis of these metavariables and proof search.

To summarize: we have described the elaboration procedure used in the
new open source interactive theorem prover \lean{} \cite{Lean}.  Our
procedure uses methods found in state-of-the-art constraints solvers,
such as nonchronological backtracking, indexing, and justification
tracking. We have also described how coercions, type classes and
ad-hoc polymorphism can be smoothly integrated in our framework using
choice constraints. Our procedure has been tested with the development
of more than 65k lines of \lean{}'s formal library, and the experience
has shown that it provides powerful and effective support for the
formalization process.

%Obvious directions for future work involve establishing the soundness
%of the elaboration process, and investigating termination and
%(semi-)completeness conditions.

\bibliographystyle{abbrv}
\bibliography{elab}

\end{document}